\documentstyle[12pt,epsf]{elsart}

\begin{document}

\begin{frontmatter}

\title{Analytical approximations of the Lindhard equations describing radiation effects}
\author[iftm]{S.Lazanu} and
\author[univ]{I.Lazanu} 

\address[iftm]{National Institute for Materials Physics, POBox MG-7,
Bucharest-Magurele, Romania, e-mail:lazanu@alpha1.infim.ro}
\address[univ]{University of Bucharest, POBox MG-11, Bucharest-Magurele,
Romania, e-mail: ilaz@scut.fizica.unibuc.ro}
%\thanks[cor]{corresponding author: fax: +40-1-4930267, e-mail:lazanu@alpha1.infim.ro}
\begin{abstract}
Starting from the general Lindhard theory describing the partition of particles energy in 
materials between ionisation and displacements, analytical approximate solutions have 
been derived, for media containing one and more atomic species, for particles identical 
and different to the medium ones. Particular cases, and the limits of these equations at very 
high energies are discussed.
\medskip

61.82.-d: Radiation effects on specific materials.\\
\medskip
\end{abstract}

\end{frontmatter}

{\it Submitted to Nuclear Instruments and Methods in Physics Research A}
\section{Introduction}

The characteristics of semiconductor devices and other crystalline materials
used in high fluences of particles are strongly affected by the effects of
radiation. In the recent years, important results have been achieved in the
radiation damage analysis, but not all the degradation mechanisms have been
completely understood up to now. After the interaction of the radiation
field with the semiconductor material, mainly two classes of degradation
effects were observed: surface damage and bulk material damage, due to the
displacement of atoms from their sites in the lattice.

If heavy particles as pions, protons, neutrons, ions, etc., produce both
surface and bulk damages, for electrons and gammas the effects are
dominantly produced at the surface. The surface effects are not a main
obstacle to semiconductor device operation, while the bulk ones have a much
greater importance. The disruption of the symmetry of the crystal,
consequence of the bulk effects, causes the formation of energy levels in
the normally forbidden region between the valence and conduction bands,
altering the material properties.

The process of partitioning the energy of the recoil nuclei (produced due
the interaction of the incident particle with the lattice site nucleus) in
new interactions processes, between electrons (ionisation) and atomic motion
(displacements) is considered in the general form, in the frame of the
Lindhard theory \cite{1} .

Up to now, there exist in the literature some calculations of the Lindhard
curves for some crystalline materials: diamond \cite{2,3}, Si \cite{4,5},
GaAs \cite{6}, InP \cite{7}, and GaP, InAs, InSb \cite{8}, in Al and some
scintillator materials \cite{9}.

The knowledge of the Lindhard energy partitioning curves is essential,
together with the detailed nuclear interaction mechanisms, in the correct
evaluation of the concentration of primary defects produced in materials
that work in intense fields of radiation. In this paper, we present the
concrete equations necessary in the analytical calculations of these curves.

\section{General hypothesis and analytical approximations of Lindhard
equations}

The general process considered in the study of the interaction between the
incident particle and the solid is the following: the particle, heavier than
the electron, with electrical charge or not, interacts with the electrons
and with the nuclei of the semiconductor lattice. It loses its energy in
several processes, which depend on the nature of the particle and on its
energy. The effect of the interaction of the incident particle with the
electrons of the target is the ionisation. The quantity characteristic for
this process is the energy loss $dE/dx$ (stopping power). The nuclear
interaction between the incident particle and the lattice nuclei produces
bulk defects. As a result of the interaction, depending on the energy and on
the nature of the incident particle, one or more light particles are formed,
and usually one (or more) heavy recoil nuclei. The nucleus has charge and
mass numbers lower or equal with that of the medium. After this interaction
process, the recoil nucleus or nuclei are displaced from the lattice
positions in interstitials. Then, the primary knock-on nucleus, if its
energy is large enough, can produce the displacement of a new nucleus, and
the process continues as long as the energy of the colliding nucleus is
higher than the threshold for atomic displacements. This phenomenon can be
regarded as a cascade process. We denote by primary displacements all the
displacements produced as a result of primary interactions, without any
further rearrangement of the vacancies and interstitials. The physical
quantity characterising the process is the concentration of primary defects
(or related quantities, for example the non ionising energy loss) produced
per unit of fluence of the incident particles.

In all the subsequent discussions, the primary recoil will be the particle
who's energy partition is to be calculated. As specified before, the primary
recoil is either a nucleus of the medium, or a nucleus with a lower mass and
charge numbers. As a consequence, for each medium a whole family of curves
can be obtained. Also these Lindhard curves can be directly used in the
evaluation of the damage produced in materials by ion beams, if the energy
of particles in the beam is identified with recoils energy.

The incident particle has an initial energy $E$, and, due to the interaction
with the target during the slowing down, this energy is transferred, on the
one side to atoms (the quantity $E_1$), and on the other side to electrons
(the quantity $E_2$). Obviously:

\begin{equation}
\label{eq.1} E=E_1+E_2 
\end{equation}

In accord with \cite{1}, the equation satisfied by $E_1$, as a function of
the energy $E$, is:
\begin{eqnarray}
\label{eq.2}
&\int d\sigma _{n,e}\left[ E_1\left( E-T_n-\sum_i T_{ei}\right)
-E_1\left( E\right)\right. \nonumber\\
&\left.+E_1\left( T_n-U\right) +\sum_i E_{1e}\left(
T_{ei}-U_i\right) \right] =0
\end{eqnarray}
where $d\sigma _{n,e}$ are the differential cross sections corresponding to
particle scattering on nucleus and electrons, $T_n\left( T_{ei}\right) $
represent the energies transferred to nucleus (and respectively to
electrons), $U$ is the energy wasted in disrupting the atomic binding, and $%
U_i$ are the corresponding ionisation energies, with summation over all
electrons.

The equation takes into consideration all the steps of the interaction until
the transferred energy is lower than the threshold energy for displacements.
In this general form, the equation is practically impossible to be solved.
It has been simplified \cite{1}, using some physical approximations: the
electrons do not produce recoil nuclei with appreciable energy, so that the
function $E_{1e}$ (for electrons), can be obtained separately; the binding
energy of the atom in the lattice can be neglected; the energy transferred
to electrons and respectively to nuclei is small in respect to the particle
energy, and electronic and nuclear collisions can be separated. More, the
energy transferred from electrons to nuclei is negligible, $E_{1e}=0$.

Using the up - mentioned approximations, the equation for the function $E_1$
becomes:

\begin{equation}
\label{eq.3} 
\left( S_n+S_e\right) E_1^{\prime }=\int^E_{T=0} E_1( T) 
\frac{d\sigma _n}{dT}dT 
\end{equation}

where $S_{n,e}=\int T_{n,e}d\sigma _{n,e}$ represent the stopping cross
sections for nuclei and electrons. The boundary condition for the equation 
\ref{eq.3} is: $E_1\left( E\right) /E\rightarrow 1$ for $E\rightarrow 0$ .

The parameter $\xi \left( E\right) $, defined as:

\begin{equation}
\label{eq.4} \ \xi \left( E\right) =\frac{S_e}{S_n} 
\end{equation}

represents a measure of the division of energy dissipation into electronic
and atomic motions. For the simplest case, of a medium consisting of only
one atomic species, and of the particle - primary recoil identical to the
particles of the medium ( $Z_{part}=Z_{med}=Z$, and $A_{part}=A_{med}=A$),
and for an electronic cross section $S_e\propto E^{1/2}$ and with $S_n$
derived from a Thomas-Fermi potential, it was shown \cite{1} that for $\xi
\left( E\right) $ there exist roughly three distinct energy regions. In the
first one, the nuclear stopping is dominating, and relatively little energy
goes into electronic motion; in the second region, the nuclear stopping
start decreasing, while the electronic one increases as $E^{1/2}$, so the
quantity $\xi \left( E\right) $ increases rapidly, and the fraction of the
energy that goes into electronic motion increases correspondingly; in the
third region, the electronic stopping starts decreasing, while $\xi \left(
E\right) $, though still increasing, approaches an asymptotic value.

It is convenient to look for analytical approximate solutions. The simplest
case to treat mathematically is that of a power low potential, $V\left(
r\right) \propto r^{-s}$, corresponding to nuclear scattering. In this case,
the equation \ref{eq.2} can be replaced with a differential equation whose
solution is the hypergeometric function. In the first region an potential
represents a good approximation. The solution is:

\begin{equation}
\label{eq.5} E_1\left( E\right) \ \propto E_c\left\{ -12+6\left[ 1+2\left(
E_c/E\right) ^{1/2}\right] \cdot \log \left( 1+\left( E_c/E\right)
^{1/2}\right) \right\} 
\end{equation}

where $E_c\propto Z\cdot A$ is the upper limit of the first energy region.
In the limit $E/E_c\ll 1$, the solution for $E_1\left( E\right) $ can be
expressed even simpler, as a power series in $\left( E/E_c\right) ^{1/2}$.
The solution described by equation \ref{eq.5} is a fast increasing one on
energy.

In the second region, the cross section could be modelled using the
Rutherford scattering formula, i.e. using $s=4/3$. The analytical formula
derived for $E_1$ \cite{1} is:

\begin{eqnarray}
\label{eq.6}
E_1\left( E\right) &\propto C_1E_b\left\{ 1-\frac 1{4\sqrt{2}
}\xi ^{-1/4}\log \frac{\xi ^{1/2}+\sqrt{2}\xi ^{1/4}+1}{\xi ^{1/2}-\sqrt{2}
\xi ^{1/4}+1}\right.\nonumber \\
&\left.-\frac {1}{2\sqrt{2}}\xi ^{-1/4}\arctan \frac{\sqrt{2}\xi ^{1/4}}{
1-\xi ^{1/2}}\right\} +C_2\xi ^{-1/4}E_b^{-1/4} 
\end{eqnarray}
where $\xi \left( E\right) $ defined by eq. \ref{eq.4} is given by $E/E_b$, $%
E_b$ being the energy at which the two stopping cross sections are equal; $%
E_b$ has the same dependence on $A$ and $Z$ as $E_c$, $C_1$ is of the order
of unity, and has the approximate expression:

\begin{equation}
\label{eq.7}C_1\simeq \frac{12}{x^2}+6\frac{\left( x+1\right) \left(
x-2\right) }{x^2}\cdot \log \left( x+1\right) 
\end{equation}
with $x\equiv E_b/E_c$ and $C_2$ is small, usually negative. The function $%
E_1\left( E\right) $ must be continuous at the boundary of the two regions.
In this region, $E_1$ has a much slower increasing slope on $E$, approaching
the plateau corresponding to the third region, and given by the asymptotic
limit.

For particles that do not belong to the medium, but for unielement media,
the problem can be solved in principle starting from the solution
corresponding to particles identical with the medium ones, obtained from
eqs. \ref{eq.5} and \ref{eq.6}, and denoted by $\epsilon \left( E\right) $.
So, the equation who's solution is of interest is:

\begin{equation}
\label{eq.8} E_1^{\prime }\left( E\right) \cdot S_{e,part}=\int d\sigma
_{part}\left[ \epsilon \left( E-T\right) -\epsilon \left( E\right) +\epsilon
\left( T\right) \right] \ 
\end{equation}

where is the electronic stopping cross section for the ion $Z_{part}$ in the
medium $Z_{med}$, and $d\sigma _{part}$ is the differential cross section
for an elastic scattering of the particle on the atom of the medium, with
the stopping cross section $S_{n,part}$. In this situation, the division of
the energy interval into three regions does not hold more, because, in place
of $S_e$ and $S_n$ we have now $S_{e,part}$ and $S_{e,med}$, , and ($%
S_{n,part}$, $S_{n,med}$ ) respectively. The case of media containing more
than one element is more complicated, and is very difficult to be treated
rigorously. A first approximation for the solution of the problem of the
energy partition of a particle in a multielement medium is to solve
separately for each component, and to use the average weight Bragg
additivity. This method has been used in the case of some binary
semiconductors \cite{7,8}.

\section{Particular cases and discussion}

In the case of a medium consisting of only one atomic species, but for
particles different from the medium, the eqs. \ref{eq.5} and \ref{eq.6}
could be still used, as a first approximation, instead of eq. \ref{eq.7},
redefining the quantities $E_c$ and $E_b$. An important distinction is to be
made between particles lighter then the medium, and heavier than the medium.
While $E_b$ has the same dependence on the characteristics of the particle
and medium in both cases:

\begin{equation}
\label{eq.9} E_b\simeq 5.2\cdot 10^{-4}\frac{\left(
Z_{part}^{2/3}+Z_{med}^{2/3}\right) ^2 }{Z_{part}^{1/3}}\frac{A_{part}^3}{%
\left( A_{part}+A_{med}\right) ^2} 
\end{equation}

(in MeV) $E_c$ has different forms.

For $Z_{part}<Z_{med}$ and $A_{part}<A_{med}$:

\begin{equation}
\label{eq.10} E_c\simeq 5\cdot 10^{-4}\frac{\left(
Z_{part}^{2/3}+Z_{med}^{2/3}\right) ^2}{Z_{part}^{1/3}}\frac{A_{part}^3}{%
\left( A_{part}+A_{med}\right) ^2} 
\end{equation}

while for $Z_{part}>Z_{med}$ and $A_{part}>A_{med}$:

\begin{equation}
\label{eq.11} E_c\simeq 1.25\cdot 10^{-4}Z_{med}\frac{\left(
A_{part}+A_{med}\right) ^2}{A_{part}} 
\end{equation}

For a given medium, consisting of only one atomic species, the family of
curves characterising the dependence of the energy channelled into
displacements, as a function of the recoils energy, and having the mass and
charge numbers ($A_{part}\leq A_{med}$ and $Z_{part}\leq Z_{med}$) as
parameters, has the following characteristics:

- The maximum energy transferred into displacements corresponds to particles
identical to the medium ones.

- All curves start, at low energies, from the same curve; they have at low
energies identical values of the energy spent in displacements, independent
on the charge and mass number of the recoil, and, roughly, an dependence.

- At higher energies, the curves start to detach from this main branch. This
happens at lower energies if their charge and mass numbers are smaller.
Then, the curves present a smooth increase with the energy. This means that
at high enough energy of the incident particle, the increase of its energy
determines, mainly, an increase of the ionisation loss. The asymptotic limit
($E_p$) of the equation \ref{eq.6} depends on the characteristics of the
particle and of the medium as:

\begin{equation}
\label{eq.12} \frac{\left( Z_{part}^{2/3}+Z_{med}^{2/3}\right) ^2}{%
Z_{part}^{1/3}}\frac{A_{part}^3}{\left( A_{part}+A_{med}\right) ^2} 
\end{equation}

In the particular case of a particle belonging to the medium, the value of
this energy is proportional to the product of the mass and charge numbers;
if, more, $Z\simeq A/2$ , as is the case of C in diamond, Si in silicon and
Ge in germanium, an $A^{\simeq 2}$ dependence is obtained for $E_p$.

\begin{figure}[tbp]
\epsfysize=8cm \epsffile[102 102 510 750]{fig1.ps hoffset=40 voffset=-110
hscale=230 vscale=230}
\caption{The displacement energy versus the particle (recoil) energy for C in diamond 
(continuous line), Si in silicon (dotted line), and respectively Ge in germanium (dashed-
dotted line). The asymptotic limits of the displacement energies are also represented. The 
characteristic energies   for the curves are indicated with arrows.}
\label{fig.1}
\end{figure}

In Figure \ref{fig.1}, the Lindhard curves for C in diamond, Si in silicon
ad Ge in germanium are presented. The values of corresponding to the
boundary between the first and the second energy regions for all three
curves are indicated by arrows. The asymptotic limits are also shown at the
highest limit of the abscissa. In fact, the solution given by equation \ref
{eq.6} is the most important in the evaluation of the bulk damage produced
in materials, in accelerator applications. For example, the standard physics
programme at LHC is based on an integrated luminosity of 5$\cdot 10^5$ pb$%
^{-1}$, which corresponds to year of operation, for an annual operation time
of 1.9$\cdot 10^7$ s. The charged hadrons are produced by the primary
interaction proton-proton at 7 GeV, while neutrons are albedo particles. The
irradiation background is continuous. The charged hadrons spectra ($\pi
^{\pm },$ $K^{\pm },$ $p,\stackrel{\_}{p}$) simulated for various positions
inside the tracking cavity of the CMS \cite{10} suggest that all hadrons
have their kinetic energies between 10$^{-2}$ - 20 GeV. Corresponding to
these kinetic energies, if only elastic interactions with the detector
medium (diamond, silicon, germanium) are considered, the recoil nuclei have
energies higher than some MeV, values in the saturation region of the
Lindhard curves. The case of binary media has been treated, in a first
approximation, as specified before, considering separately the two
components, and then weighting with their number in the molecule. If the two
components are adjoining each other in the periodic table, as is the case
of, e.g., GaAs and InSb, then there is no significant difference in the
Lindhard partition energy curves, for element placed between the two
elements of the compound. In Figure \ref{fig.2}, the Lindhard curves for As
\begin{figure}[tbp]
\epsfysize=8cm \epsffile[102 102 510 750]{fig2.ps hoffset=30 voffset=-110
hscale=230 vscale=230}
\caption{The displacement energy versus particle (recoil) energy for two sets of elements near 
one another in the periodic table, that form two binary   compounds. The first one indicates 
the particle, the second one the medium. The asymptotic limits (plateau) of the displacement 
energies are also represented.}
\label{fig.2}
\end{figure}
in Ga, Ga in As, Sb in In and In in Sb are given. For all of them, the
asymptotic limit is also indicated. The curves for Ga and As could be
compared with the one corresponding to Ge in germanium, presented in Figure 
\ref{fig.1}. If the two components are far each other in the periodic table,
then there are, in principle, two families of curves, one corresponding to
\begin{figure}[tbp]
\epsfysize=8cm \epsffile[102 102 510 750]{fig3.ps hoffset=30 voffset=-110
hscale=230 vscale=230}
\caption{The displacement energy versus particle (recoil) energy for two sets of elements far 
one another in the periodic table, that form two binary compounds. The first one indicates 
the particle, the second one the medium. The asymptotic limits of the displacement energies 
are also represented.}
\label{fig.3}
\end{figure}
the heaviest element, the other to the lightest one, separated by a gap. As
an illustration, in Figure \ref{fig.3} the results for P in In and In in P,
as well as those for Ga in P and P in Ga are presented. These are useful in
the evaluation of radiation defects in the semiconductor compounds GaP and
InP. The asymptotic limit of the energy used for displacements is also
represented.

\section{Summary}

Recent development of accelerator machines conduces to the necessity of the
study of radiation effects in a multitude of materials, in intense fields of
a large variety of particles, in a high range of energy and at different
fluences. As essential factor in the evaluation of radiation effects is the
knowledge of the energy partition of slowing down particles, between
ionisation and displacements in the lattice of the target. The Lindhard
theory gives a detailed description of the calculation procedure for
particles identical to the medium and for media consisting of one atomic
species, and proposes some approximations for other situations. In the
present article, we applied approximate methods for the calculation of the
Lindhard curves for particles that do not belong to the medium, both for
unielement and compound media. As the asymptotic value of these curves is of
much interest for practical applications, the dependence of this value on
the characteristics of the particle and medium is given.

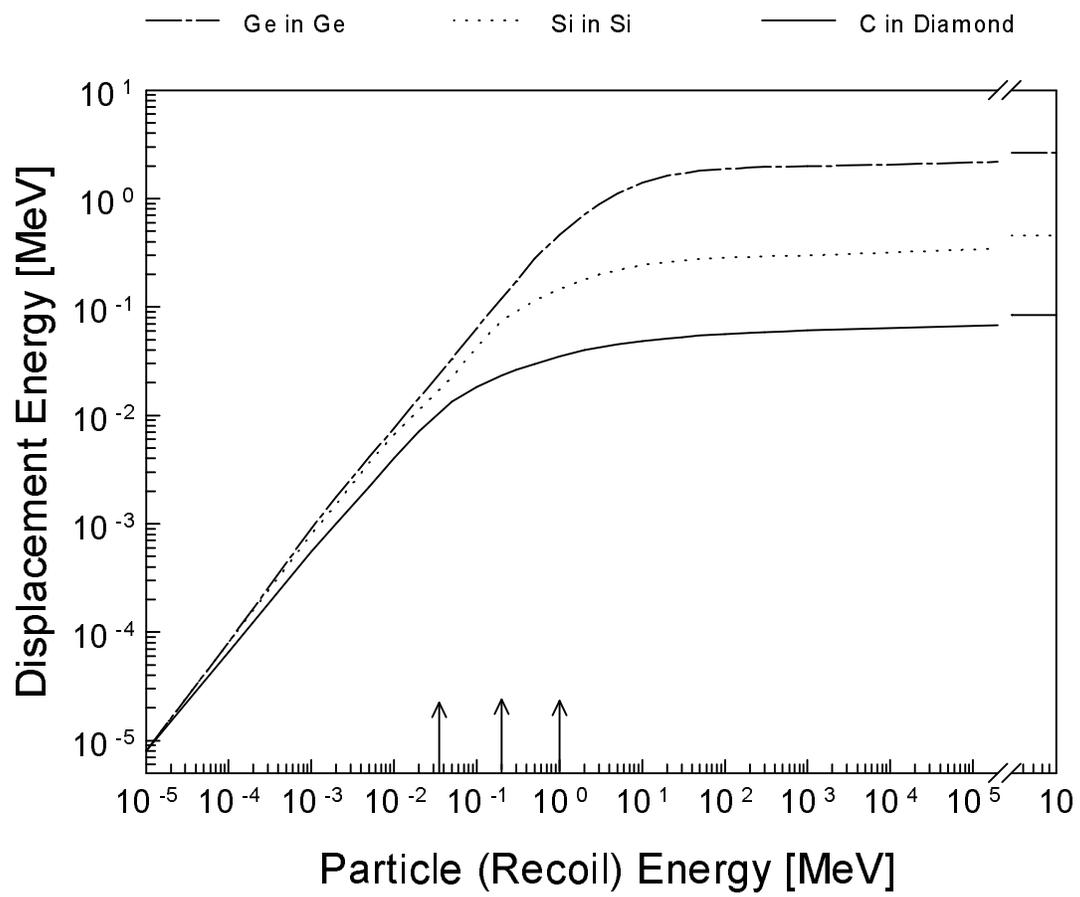
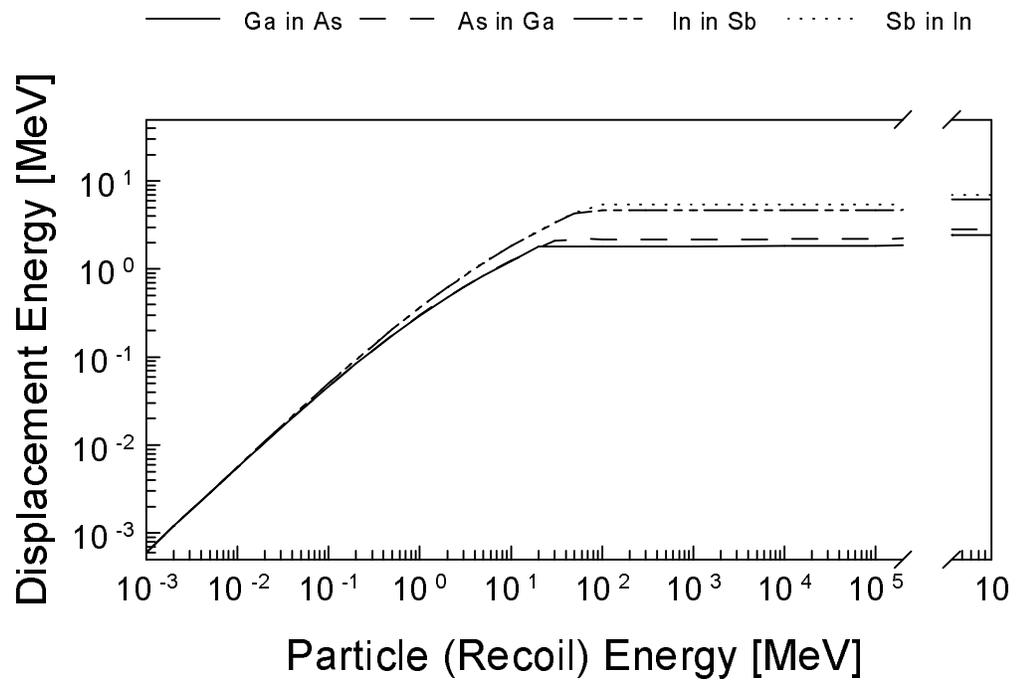
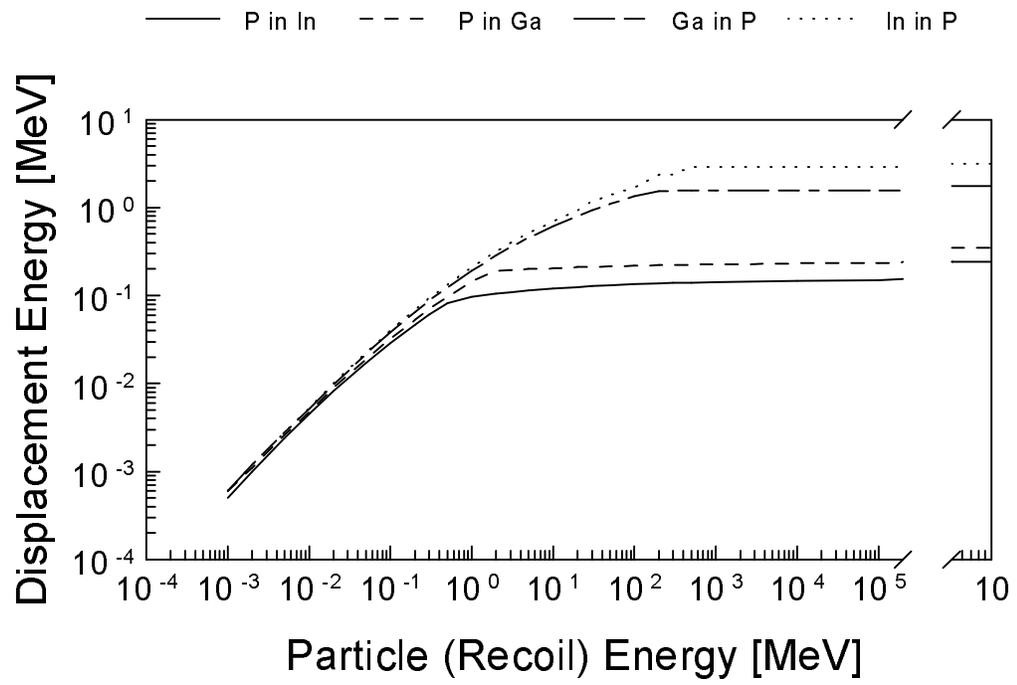
\end{document}